\documentclass[11pt,a4paper]{article}
\usepackage[utf8]{inputenc} 

\usepackage{amsmath,amssymb,amscd,amsfonts,mathabx, textcomp}
\usepackage{amsthm}
\usepackage{floatflt}

\begin{document}

\title{\bf Disformal transformations\\ in modified teleparallel gravity}

\author{Alexey Golovnev$^{1}$ and Mar\'ia Jos\'e Guzm\'an$^{2}$\\ \\
$^{1}$ \it Centre for Theoretical Physics, The British University in Egypt,\\ \it 11837 El Sherouk City, Egypt;\\  \small agolovnev@yandex.ru\\ \\
$^{2}$ \it Departamento de F\'isica y Astronom\'ia, Facultad de Ciencias, Universidad de La Serena,\\ \it Av. Juan Cisternas 1200, 1720236 La Serena, Chile;\\ \small  maria.j.guzman.m@gmail.com}
\date{}

\maketitle

\abstract{In this work, we explore disformal transformations in the context of the teleparallel equivalent of general relativity and modified teleparallel gravity. We present explicit formulas in components for disformal transformations of the main geometric objects in these theories such as torsion tensor, torsion vector and contortion. Most importantly, we consider the boundary term which distinguishes the torsion scalar from the Ricci scalar. With that we show for $f(T)$ gravity that disformal transformations from the Jordan frame representation are unable to straightforwardly remove local Lorentz breaking terms that characterize it. However, we have shown that disformal transformations have interesting properties, which can be useful for future applications in scalar-torsion gravity models, among others.
}

\section{Introduction}

Modifications to Einstein's general relativity are nowadays a well motivated field of research, due to the riddles of dark sectors in cosmology such as dark energy and dark matter. Moreover, the~need for an early accelerated expansion in the inflationary paradigm, as~well as infamous obstacles in the quest for quantum gravity, can also be seen as indications of new physics beyond general~relativity.

One of the possible paths to deform general relativity lies in modification of the fundamental geometric objects that describe the gravitational field. It is known that there exist at least two alternative equivalent descriptions of general relativity based on torsion or non-metricity of spacetime. They are called teleparallel equivalent of general relativity (TEGR) \cite{Aldrovandi:2013wha,Golovnev:2018red} and symmetric teleparallel gravity (STEGR)\cite{Adak:2004uh,Adak:2005cd}, respectively. The~Lagrangians are proportional to the torsion scalar $\mathbb T$ and the non-metricity scalar $\mathbb Q$, which differ from each other and from the Ricci scalar $R$ only by  boundary terms, giving rise to equivalent equations of motion. However, their modifications, for~example $f(\mathbb T)$~\cite{Ferraro:2006jd} and $f(\mathbb Q)$ \cite{Jimenez:2019ovq}, yield genuinely different gravity~models.

Even the simplest modified models of this sort, such as $f(\mathbb T)$, are very non-trivial and present challenges for understanding the nature of their degrees of freedom (see Section~\ref{sec:sumdof} for a summary). Note that, in other cases, clever changes of variables have proven to be very useful for investigating the structure of theories at hand. For~example, in~$f(R)$ gravity, conformal transformations allow to establish equivalence with scalar-tensor models. At~the same time, in~$f(\mathbb T)$ they fall short of giving a clearcut result. Then, it feels natural to try a more general class of transformations, the~disformal~ones.

Disformal transformations have been considered by Bekenstein as a deformation of the metric tensor that defines a gravitational theory supplied by two geometries, one for the
gravity sector and one for the matter sector. Assuming Finslerian geometry
for the matter sector, he found that in order to preserve both the weak equivalence
principle and the causal structure, it has to reduce to a Riemannian geometry,
whose metric $\tilde{g}_{\mu\nu}$ is related to the gravitational one $g_{\mu\nu}$ by
~\cite{Bekenstein1992mgm,Bekenstein:1992pj}
\begin{equation}
\tilde{g}_{\mu\nu} = \alpha^{2} g_{\mu\nu} + \beta \partial_{\mu} \phi \partial_{\nu} \phi.
\end{equation}

This is a non-trivial generalization of a conformal transformation, in~the sense that the original space-time $g_{\mu\nu}$ is not only isotropically stretched but also considers anisotropic variation in the directions in which the scalar field varies. 
The causal structure of the transformed metric depends on the coefficients $\alpha,\beta$, which have to be restricted at least by requiring that the metric is non-degenerate. One of the interesting applications of disformal transformations to gravity is their use for mapping gravitational theories with second-order field equations to higher derivative theories that still avoid Ostrogradsky's instability~\cite{Horndeski:1974wa}.

This sort of transformation has been scarcely discussed in the context of the tetrad \mbox{formalism~\cite{Bittencourt:2015ypa,Hohmann:2019gmt}.} 
We are interested in developing disformal transformations in the teleparallel equivalent of general relativity framework, and~apply it to modified teleparallel gravity in order to better understand the issue of the extra degrees of freedom that these theories exhibit. For~this, we~start with a brief introduction to teleparallel gravity in Section~\ref{sec:tegr}. We introduce modifications of gravity based on the teleparallel formalism in Section~\ref{sec:fTandDof}, together with an introduction to the issue of the degrees of freedom. The~basics of disformal transformations in tetrad formalism and transformations of the main geometrical objects are presented in Section~\ref{sec:disformaltetrad}. The~main results of this paper are presented in Section~\ref{sec:disformalfT}, where we look for an interpretation of the extra d.o.f. of $f(\mathbb T)$ gravity in terms of a disformal transformation parameterized by a scalar field. Finally, we devote Section~\ref{sec:concl} to the~conclusions.

\section{Teleparallel gravity}
\label{sec:tegr}

The spacetime manifold in which gravitational phenomena take place is commonly described by two geometrical structures: a base of orthonormal vectors (tetrad or {\it vierbein}) and a spin~connection. The~first one leads to the definition of metric, while the second one provides the rule for parallel~transport. The~basis of tetrads $\mathbf{e}_a$ is defined in the tangent space $T_p(M)$ of a 4-dimensional manifold $M$, and~the corresponding co-tetrad basis $\mathbf{E}^a$ resides in the co-tangent space $T^{\star}_p(M)$. The~Latin indices $a,b,\ldots =0,\ldots,3$ will denote (co-)tangent space indices, and~the Greek indices \mbox{$\mu,\nu,\ldots = 0, \ldots, 3$} are spacetime ones. When expanded in a coordinate basis as $\mathbf{e}_a=e^{\mu}_{a} \partial_{\mu}$ and $\mathbf{E}^{a}=E^{a}_{\mu} dx^{\mu}$, the~components satisfy completeness relations $E^{a}_{\mu} e^{\mu}_b = \delta^{a}_{b}$ and $E^{a}_{\mu} e^{\nu}_a = \delta^{\nu}_{\mu}$. We recover the usual metric of general relativity by declaring the norm of tetrads in terms of the Minkowski metric, that is
\begin{equation}
e^{\mu}_a e^{\nu}_b g_{\mu\nu} = \eta_{ab}, \ \ \ \ \ g_{\mu\nu} = E^{a}_{\mu} E^{b}_{\nu} \eta_{ab},
\label{orthonorm}
\end{equation}
where, in the second relation, we show that the metric can be recovered through the co-tetrad components. We will consider a spacetime endowed with the Weitzenb\"{o}ck connection, defined~as
\begin{equation}
\Gamma^{\alpha}_{\ \mu\nu}=e^{\alpha}_a \partial_{\mu} E^a_{\nu}.
    \label{Wconn}
\end{equation}
This is equivalent to the choice of the spin connection $\omega^{a}_{\ \mu b}=0$, and~it renders the Riemann tensor to vanish, but~leaves non-zero torsion $T^{\alpha}_{\ \mu\nu} = \Gamma^{\alpha}_{\ \mu\nu} - \Gamma^{\alpha}_{\ \nu\mu} =  e^{\alpha}_a (\partial_{\mu} E^{a}_{\nu} - \partial_{\nu} E^{a}_{\mu} )$. In~such a spacetime, the~basis vectors are parallel, transported trivially throughout the manifold, coining the term tele-parallel (parallel at a distance). 

The Weitzenb\"{o}ck connection is just a subcase of a more general family of flat connections. In~general, the condition of flatness is imposed as vanishing of the curvature  2-form. This condition can be solved by the so-called inertial connection~\cite{Aldrovandi:2013wha,Krssak:2015rqa}, which can also be viewed as covariantization of the  Weitzenb\"{o}ck case since it can be obtained by arbitrary Lorentz transformations of the~latter.

A gravitational theory that gives dynamics to the tetrad field and has equations of motion equivalent to the Einstein's equations is the teleparallel equivalent of general relativity (TEGR), whose action is written as (we adopt the $(+,-,-,-)$ signature convention, so that the Lagrangian density of the Einstein--Hilbert action reads $-\sqrt{g}R$, with~the minus sign) \cite{Aldrovandi:2013wha,Golovnev:2018red}
\begin{equation}
S = \dfrac{1}{2\kappa} \int d^{4}x\ E \ \mathbb T,
\label{Ltegr}
\end{equation}
where $\kappa=8\pi G$, $E = \text{det}(E^{a}_{\mu})$, and~\begin{equation}
\mathbb T = \dfrac{1}{4} T^{\rho}_{\hphantom{\rho} \mu\nu} T^{\hphantom{\rho} \mu\nu}_{\rho} - \dfrac{1}{2} T^{\rho}_{\hphantom{\rho} \mu\nu} T^{\mu\nu}_{\hphantom{\mu\nu} \rho} - T_{\mu} T^{\mu},
\label{Tscalar}
\end{equation}
where $T_{\mu}=T^{\rho}_{\hphantom{\rho} \mu\rho}$ is the torsion~vector.

The TEGR action is equivalent to general relativity (GR) due to the mathematical identity
\begin{equation}
R(\Gamma)=R(\mathop\Gamma\limits^{(0)})+2 \mathop\bigtriangledown\limits^{(0)}{\vphantom{\bigtriangledown}}_{\mu}T^{\mu}+\mathbb T
\end{equation} 
between the curvatures of a metric-compatible torsionful connection $\Gamma$ and the Levi-Civita connection $\mathop\Gamma\limits^{(0)}$. With~the zero spin connection prescription, the~curvature $R(\Gamma)$ is equal to zero, and~the TEGR action given in terms of the torsion scalar is equivalent to the Einstein--Hilbert one modulo the surface term of the torsion vector divergence, this is
\begin{equation}
- E \mathbb T = E R(\mathop\Gamma\limits^{(0)}) + 2 E  \mathop\bigtriangledown\limits^{(0)}{\vphantom{\bigtriangledown}}_{\mu}T^{\mu}.
\label{equiv}
\end{equation}

Given the Weitzenb\"{o}ck choice of the connection, the~surface term is not invariant under the local Lorentz transformations (LLT) $E^{a}_{\mu} \rightarrow E^{a \prime}_{\mu} = \Lambda^{a\prime}_{a}(x) E^{a}_{\mu}$. This fact is not important once we plug this expression into the TEGR action, since it is integrated out and it gives Lorentz invariant equations of motion fully equivalent to GR. However, it will be a crucial point when taking non-linear extensions of this~model.

\section{Modified teleparallel gravity and degrees of freedom}
\label{sec:fTandDof}

\subsection{Generalized teleparallel gravity}

The Lagrangian \eqref{Ltegr} is a suitable starting point for building modifications to Einstein's gravity. This is because the Lagrangian is quadratic in first-order derivatives of the tetrad field; therefore, any non-linear modification of it would yield equations of motion with only second-order derivatives. This outcome is not easy to get once we adopt the task of modifying gravity, as~it is quite common to get fourth-order equations of motion when such modifications are implemented. This theoretical motivation led to the introduction of the $f(\mathbb T)$ gravity theories with the action~\cite{Ferraro:2006jd,Bengochea:2008gz}
\begin{equation}
S = \dfrac{1}{2\kappa} \int d^{4}x \ E\ f(\mathbb T ).
\end{equation}
The associated equations of motion are indeed purely second order in derivatives of the tetrad~field,~namely
\begin{equation}
4  S_{a}^{\hphantom{a} \mu\nu}  f^{\prime\prime}(\mathbb T)\partial_{\mu} \mathbb T + 4 \left[ e^{\lambda}_a T^{\rho}_{\hphantom{\rho} \mu\lambda} S_{\rho}^{\hphantom{\rho} \mu\nu} + e \partial_{\mu}(E e^{\lambda}_a S_{\lambda}^{\hphantom{\rho} \mu\nu} ) \right] f^{\prime}(\mathbb T) - e^{\nu}_a f(\mathbb T) = -2\kappa e^{\lambda}_a \mathcal{T}_{\lambda}^{\nu},
\end{equation}
where $\mathcal{T}$ refers to the usual energy--momentum tensor of any type of matter which couples to the~metric. These equations are obtained by taking the variation with respect to $E^{a}_{\nu}$, and~we can easily see their second-order character. 
In addition to this, the~theory has interesting applications in cosmology, a~field in which it has been extensively used for getting early and late accelerated phases without introducing additional~fields. 

When working in the pure tetrad formalism, the~$f(\mathbb T)$ theory, as~well as most other modified teleparallel theories, exhibits violation of the local Lorentz invariance. In~particular, for~$f(\mathbb T)$, the reason can be traced to Equation \eqref{equiv}. Since the torsion scalar changes by a surface term under a local Lorentz transformation, the~non-linear function of it will transform as
\begin{equation}
f(\mathbb T) \longrightarrow f(\mathbb T + \text{surface term}).
\end{equation}
Therefore, the~former surface term can no longer be integrated out, and~it gives rise to a theory sensitive to local Lorentz~transformations.

A different angle regarding this issue can be obtained in the Jordan frame representation of $f(\mathbb T)$ gravity. For~this, we rewrite the action with the help of an auxiliary scalar field $\phi$ as
\begin{equation}
S = \dfrac{1}{2\kappa} \int d^4 x  E \left(\vphantom{\int}f(\phi)-f^{\prime}(\phi)\cdot({\mathbb T}-\phi)\right) ,
\label{Jfr}
\end{equation}
for which the equations of motion are the same as for the original model, and~$\phi=\mathbb T$. Another option is to choose the scalar field as $\phi=f^{\prime}(\mathbb T)$ (related to the previous one via an obvious field redefinition), so that
\begin{equation}
\label{JfrL}
S  = - \dfrac{1}{2\kappa} \int d^4 x \ E \ (\phi \mathbb T - V(\phi) ),
\end{equation}
and it can be viewed as a Legendre transform $V(\phi) = \mathbb T \phi - f(\mathbb T)$, and~$\mathbb T=V^{\prime}(\phi)$. In this formulation, one can relate the symmetry breaking with the scalar field $\phi$ changing in space and time, since now the former surface term produces effects via derivatives acting upon $f^{\prime}(\phi)$ or $\phi$ at an attempt to integrate by parts. This resonates well with the fact that the Lorentz-violating terms in $f(\mathbb T)$ equations of motion are proportional to $\partial \mathbb T$, and~the scalar $\phi$ is set to be given by $\mathbb T$.

A natural next step is to try removing the extra factor in front of $\mathbb T$ via a conformal~transformation.
A known fact is that, in teleparallel gravity, this approach falls short of success. A~would-be Einstein frame (where the factor in front of $\mathbb T$ is the unity) results, with another Lorentz-violating term~\cite{Yang:2010ji,Wright:2016ayu}.
The action that results after this procedure is
\begin{equation}
S = \dfrac{1}{2\kappa} \int d^{4}x\ E\ \left( \hat{\mathbb T} - \dfrac{2}{\sqrt{3}} \hat{T}^{\mu}\partial_{\mu} \psi  - \dfrac{1}{2}\hat{g}^{\mu\nu} \partial_{\mu}\psi \partial_{\nu}\psi - U(\psi) \right),
\label{EinsConf}
\end{equation}
where hatted quantities refer to conformally transformed~ones. This frame comes together with the presence of a scalar field with an unusual coupling $\hat{T}^{\mu}\partial_{\mu}\psi$, which spoils local Lorentz invariance. This formulation is very suggestive of the presence of new degree(s) of freedom, and~also has the worrisome feature that the kinetic term of the $\psi$ field is of the ghostly sign. However, one should not forget that there is an intricate kinetic mixing of the scalar field with the tetrad components via the Lorentz-breaking term. Therefore, it is not easy to draw definitive conclusions from this form of the~action.

It is, of course, not surprising that we could not cleanly reveal these properties via a conformal~transformation. It should not be possible to get rid of Lorentz-violating terms by an isotropic field~redefinition. However, we could definitely try to explicitly relate those effects to some different kinds of fields. Like, for~example, preferred frame effects can come in the disguise of a vector field in the Einstein--Aether model~\cite{Eling:2004dk}. Furthermore, since, in our case, the origin of violation seems to be related with the gradient of the scalar, we can try a more complex transformation, which  contains a preferred direction---a~disformal transformation. The~assertion we would like to test is whether it can be used to parameterize the new dynamical mode of $f(\mathbb T)$ \cite{Ferraro:2018tpu}, and~if it is possible to remove Lorentz-violating terms in a teleparallel Einstein frame. Before~doing so, let us briefly summarize the state-of-the-art concerning the issue of  the degrees of~freedom.

\subsection{On the extra degree(s) of freedom of f(T) gravity}
\label{sec:sumdof}

The issue of the nature and the number of the extra degree(s) of freedom of $f(\mathbb T)$ gravity and other modified teleparallel theories is still not settled, despite intense debate in recent years~\cite{Ferraro:2018tpu,Jimenez:2019tkx,Jimenez:2019ghw,Koivisto:2018loq,Raatikainen:2019qey}. Here, we will briefly summarize some points that we feel are important to consider in a~discussion:
\begin{itemize}
\item {\it Lorentz invariance}: It is believed by some authors~\cite{Aldrovandi:2013wha,Krssak:2015oua,Golovnev:2017dox,Hohmann:2018rwf,Krssak:2018ywd} that the loss of local  Lorentz invariance could be restored by  dropping the assumption of the Weitzenb\"{o}ck connection and resorting to a more general one, inertial connection. However, see the Ref.~\cite{Bejarano:2019fii} for another viewpoint, especially when it comes to modified teleparallel gravities.
\item {\it Weak gravity}: It has been shown that there are no new propagating gravitational modes in the linearized approximation around Minkowski spacetime with the trivial choice of the tetrad~\cite{ Bamba:2013ooa,Cai:2018rzd}. This follows from a simple observation that $f(\mathbb T)$ gravity reduces to TEGR in this limit~\cite{Hohmann:2018jso}, and~in particular, the local Lorentz invariance is restored.
\item {\it Cosmological perturbations}: Extra propagating modes also do not appear in the linear cosmological perturbations around a spatially flat FRW universe with the standard choice of the background tetrad~\cite{Chen:2010va,Li:2011wu,Izumi:2012qj,Golovnev:2018wbh,Toporensky:2019mlg}. When a dynamical scalar mode is seen in the perturbations~\cite{Izumi:2012qj}, it actually comes from an additional scalar field added as a matter field to the model.
\item {\it Hamiltonian formalism}: Probably, the~first attempt to count the number of degrees of freedom in the Hamiltonian formalism of $f(\mathbb T)$ gravity can be found in the Ref.~\cite{Li:2011rn}, where the authors claim that the theory has $n-1$ extra degrees of freedom in $n$ spacetime dimensions.  Later work finds only one extra degree of freedom in any dimension~\cite{Ferraro:2018tpu,Ferraro:2018axk}. It is suggested that the extra mode could be related with the proper parallelization of spacetime. For~other kinds of modified teleparallel gravities, the~consideration of this mode is still pending~\cite{Blixt:2018znp}.
\item {\it Conformal transformations}: We proved above that a clean interpretation of the extra degree(s) of freedom of $f(\mathbb T)$ gravity cannot be achieved with the use of conformal transformations. This result was first obtained in~\cite{Yang:2010ji}, and~some more details and refinements, and~also extensions have been given in~\cite{Wright:2016ayu}.
\item {\it Claims of acausality}: Assuming the presence of extra modes, the~cosmological results indicate a strong coupling regime for them, which is bad news for predictability. Also, acausality claims for $f(\mathbb T)$ gravity can be found in the Ref.~\cite{Ong:2013qja} through the method of characteristics; however, they use the number of degrees of freedom calculated in the Ref.~\cite{Li:2011rn} to draw some of their conclusions, which are currently in doubt.
\item {\it Galaxy rotation curves}: An indirect indication for extra degree(s) of freedom can be found in~\cite{Finch:2018gkh}, where they use a model with $f(\mathbb T)=\mathbb T+\alpha \mathbb T^n$ to reproduce galactic rotation curves. When the function approaches GR/TEGR ($n=1$), the~curves are different to those obtained in GR, indicating the presence of some extra mode. 
\item {\it Remnant symmetry}: Even though the full Lorentz group is no longer a symmetry of $f(\mathbb T)$ gravity, there still exist subgroups of remnant symmetries which have been studied in the Ref.~\cite{Ferraro:2014owa}. Obviously, a good understanding of symmetries is an important key to understanding the degrees of freedom.
\item {\it Null tetrads}: Last but not least, null tetrads can be used to find solutions with $\mathbb T=0$ (or any other constant instead of zero) \cite{Bejarano:2014bca,Nashed:2014yea,Bejarano:2017akj}, which could be useful for exhibiting the extra d.o.f.~\cite{Guzman:2019oth}, even though, at the background level, these solutions are not different from general relativistic ones, up~to rescaling of fundamental constants.
\end{itemize}

Based on all the points raised above, and~focusing particularly on isolating the extra d.o.f. in $f(\mathbb T)$ gravity through some transformation involving a scalar field, it comes naturally to ask whether an extended kind of transformation, the~so called disformal transformations, can give some clue about the physical interpretation of the extra d.o.f.. This work is intended to fill this gap in current research by presenting basic calculations in (modified) teleparallel theories of gravity and possible interpretations and~applications.

\section{Disformal transformations in the teleparallel\\ framework}
\label{sec:disformaltetrad}

\subsection{Generalities on disformal transformation of the tetrad field}

We will consider disformal transformations of the tetrad field in the following general form:
\begin{equation}
{\tilde E}^a_{\mu}=C\cdot E^a_{\mu}+D\cdot (\partial^a \phi)(\partial_{\mu}\phi),
\end{equation}
where we use the standard formulas for manipulating tangent space indices, that is $\partial_a \equiv e^{\mu}_a \partial_{\mu}$ and $\partial^{a} = \eta^{ab} \partial_{b}$. The~$C$ and $D$ are scalar functions, and~for the majority of calculations, we will not need to explicitly specify their arguments, although, ultimately, we assume they depend on $\phi$ and $(\partial\phi)^2 \equiv  (\partial_{\mu}\phi)(\partial^{\mu}\phi)$.

Note that the tetrad transformation is quadratic in the field derivatives. However, it is easy to prove that the metric transforms the same way, though~with the $(\partial\phi)^2$ correction in the coefficient, see~below. That is why it would not be appropriate 
 in our case to restrict the arguments of $C$ and $D$ to the value of $\phi$ only.
In the current setting, we even could have allowed dependence on $\mathbb T$. \mbox{However, our main} purpose is to see whether it is possible to get rid of the local Lorentz-violating terms in the Lagrangian. It is hard to imagine this goal could be achieved if all coefficients in every term depend on the torsion~scalar.

We introduce the shorthand notation $\phi_{\mu} \equiv \partial_{\mu}\phi$, and~analogously, for any other scalar function, such as $C$ or $D$, and~use it whenever convenient. With~this, we can write
\begin{equation}
{\tilde E}^a_{\mu}=C E^a_{\mu}+D\phi^a \phi_{\mu},
\label{dttetr}
\end{equation}
\begin{equation}
{\tilde e}_a^{\mu}=\frac{1}{C}\left(e_a^{\mu}-\frac{D}{C+D(\partial\phi)^2}\phi_a \phi^{\mu}\right),
\label{dtitetr}
\end{equation}
for the tetrad and its inverse. Since one is the inverse of another, the~orthonormality relations ${\tilde E}^{a}_{\mu} {\tilde e}^{\mu}_b = \delta^{a}_{b}$ and ${\tilde E}^{a}_{\mu} {\tilde e}^{\nu}_a = \delta^{\nu}_{\mu}$ still~hold. 

One can easily obtain the transformation of the metric \eqref{orthonorm} and its inverse as
\begin{equation}
{\tilde g}_{\mu\nu}=C^2 g_{\mu\nu}+D(2C+D(\partial\phi)^2)\phi_{\mu}\phi_{\nu}
\end{equation}
and
\begin{equation}
{\tilde g}^{\mu\nu}=\frac{1}{C^2}\left( g^{\mu\nu}-\frac{D(2C+D(\partial\phi)^2)}{(C+D(\partial\phi)^2)^2}\phi^{\mu}\phi^{\nu}\right).
\end{equation}
A useful consequence of the last relation is that
\begin{equation}
\tilde{\phi^{\mu}}\equiv {\tilde g}^{\mu\nu}\partial_{\nu}\phi=\frac{1}{(C+D(\partial\phi)^2)^2}\phi^{\mu}    
\end{equation}
which simply states that the length of $\phi_{\mu}$ transforms according to the corresponding eigenvalue of the transformation~matrix.

{An important requirement when doing transformations in a given theory is to preserve~invertibility. For~the disformal transformations, the most obvious requirement amounts to}
$$C\neq 0, \qquad C+D(\partial\phi)^2\neq 0$$
which is the condition on the eigenvalues of the transformation matrix. If~a transformation should be continuously connected with identity, then we substitute the $\neq$ signs with the $>$ ones.

Since the transformations above contain field derivatives, they can fail being invertible even with non-zero eigenvalues, thus producing what is known under the name of mimetic gravity~\cite{Chamseddine:2013kea,Golovnev:2013jxa}. The~corresponding general conditions for disformal transformations of the metric can be found in the Ref.~\cite{Deruelle:2014zza}.

{Sometimes conditions on light cones, time directions and other properties of the metrics are also~imposed. They can be easily derived, too. However,~we will not need any of these. Now, we have all the basic building blocks for studying transformations of the main geometrical objects in teleparallel~gravity. }

\subsection{Disformal transformation of the torsion tensor}

We start with the calculation of the disformally transformed connection coefficients, defined in Equation \eqref{Wconn}. The~calculation amounts to replacing the tetrads by their tilde versions given in formulas~\eqref{dttetr} and \eqref{dtitetr}. Before~that, and~considering that  $E^{a}_{\alpha} \Gamma^{\alpha}_{\ \mu\nu} = \partial_{\mu} E^{a}_{\nu}$, we can make a useful observation for the object $\partial_{\mu} \phi^{a} $:
\begin{equation}
\partial_{\mu} \phi^{a} = \partial_{\mu} ( E^{a}_{\nu} \phi^{\nu} ) = E^{a}_{\nu} \partial_{\mu} \phi^{\nu} + \Gamma^{\nu}_{\ \mu\alpha} E^{a}_{\nu} \phi^{\alpha} = \nabla_{\mu} \phi^{a} = E^{a}_{\nu} \nabla_{\mu} \phi^{\nu}.
\end{equation}
Now, with~all these prerequisites, we get
\begin{equation}
\begin{split}
{\tilde \Gamma}^{\alpha}_{\mu\nu} & =\Gamma^{\alpha}_{\mu\nu}+\frac{C_{\mu}}{C}\delta^{\alpha}_{\nu}+\frac{D}{C+D(\partial\phi)^2}\phi^{\alpha} \bigtriangledown_{\mu}\phi_{\nu}+\frac{CD_{\mu}-DC_{\mu}}{C(C+D(\partial\phi)^2)}\phi^{\alpha}\phi_{\nu} \\ 
& +\frac{D}{C}\phi_{\nu}\bigtriangledown_{\mu}\phi^{\alpha}-\frac{D^2}{C(C+D(\partial\phi)^2)}\phi^{\alpha}\phi^{\beta}\phi_{\nu}\bigtriangledown_{\mu}\phi_{\beta}.
\end{split}
\label{disfConn}
\end{equation}
The disformally transformed torsion tensor is the antisymmetric part of this expression \eqref{disfConn}.
Note, that if we want to specify the admissible arguments, for~example, that $C=C(\phi,(\partial\phi)^2)$, we just need to use the  Leibniz rule, that is
\begin{equation}
C_{\mu} \equiv \partial_{\mu}C(\phi,(\partial\phi)^2)=\frac{\partial C}{\partial\phi}\phi_{\mu} +2\frac{\partial C}{\partial(\partial\phi)^2}\phi^{\alpha}\bigtriangledown_{\mu}\phi_{\alpha}.
\end{equation}
Another way to obtain $\tilde{T}^{\alpha}_{\hphantom{\alpha} \mu\nu} $ is to use mixed components (differential forms) representation, this is   $\tilde{T}^a_{\hphantom{\beta}\mu\nu}=\partial_{\mu}\tilde{E}^a_{\nu}-\partial_{\nu}\tilde{E}^a_{\mu}$. With~our definitions, we have
\begin{equation}
{\tilde T}^a_{\hphantom{\beta}\mu\nu}= CT^a_{\hphantom{\beta}\mu\nu} +C_{\mu}E^a_{\nu}-C_{\nu}E^a_{\mu} + \phi^a(\phi_{\nu} D_{\mu}-\phi_{\mu}D_{\nu}) +D(\phi_{\nu}\bigtriangledown_{\mu}\phi^a-\phi_{\mu}\bigtriangledown_{\nu}\phi^a).
\end{equation}
This result has also been obtained in~\cite{Hohmann:2019gmt}. However, the~expression we want is ${\tilde T}^{\alpha}_{\hphantom{\alpha}\mu\nu}={\tilde e}^{\alpha}_a {\tilde T}^a_{\hphantom{\alpha}\mu\nu}$, which is easily found to coincide with the antisymmetric part of the formula (\ref{disfConn}):
\begin{equation}
\label{disftors}
\begin{split}
{\tilde T}^{\alpha}_{\hphantom{\alpha}\mu\nu} & = T^{\alpha}_{\hphantom{\alpha}\mu\nu} - \frac{D}{C+D(\partial\phi)^2}\phi_{\beta}\phi^{\alpha}T^{\beta}_{\hphantom{\beta}\mu\nu} + \frac{C_{\mu}}{C}\delta^{\alpha}_{\nu}-\frac{C_{\nu}}{C}\delta^{\alpha}_{\mu}+\frac{(CD_{\mu}-DC_{\mu})\phi_{\nu} - (CD_{\nu}-DC_{\nu})\phi_{\mu}}{C(C+D(\partial\phi)^2)}\phi^{\alpha} \\    
& +\frac{D}{C}(\phi_{\nu}\bigtriangledown_{\mu}\phi^{\alpha}-\phi_{\mu}\bigtriangledown_{\nu}\phi^{\alpha})-\frac{D^2}{C(C+D(\partial\phi)^2)}\phi^{\alpha}\phi_{\beta}(\phi_{\nu}\bigtriangledown_{\mu}\phi^{\beta}-\phi_{\mu}\bigtriangledown_{\nu}\phi^{\beta})
\end{split}    
\end{equation}
where we have taken into account that
\begin{equation}
\bigtriangledown_{\mu}\phi_{\nu}-\bigtriangledown_{\nu}\phi_{\mu}=-T^{\beta}_{\hphantom{\beta}\mu\nu}\phi_{\beta}.
\end{equation}

\subsection{Contortion and torsion vector}

In order to find the Einstein frame for $f(T)$ gravity, a~straightforward way to proceed would be to find the transformation of the torsion scalar. It amounts to calculating
\begin{equation}
\tilde{\mathbb{T}}={\tilde T}^{a}_{\hphantom{a}\mu\nu}{\tilde T}^{b}_{\hphantom{b}\rho\sigma}\left(\frac14 \eta_{ab}{\tilde g}^{\rho\mu}{\tilde g}^{\nu\sigma}+ \frac12 {\tilde e}^{\rho}_a {\tilde e_b^{\mu} } {\tilde g}^{\nu\sigma}-{\tilde e}^{\nu}_{a}{\tilde e}^{\sigma}_{b}{\tilde g}^{\mu\rho}\right).
\label{almostprem}
\end{equation}
However, this calculation is cumbersome, and~we leave it for future work. Additionally, as~we will see later, for our purpose, it is possible to take a shortcut and calculate only the transformation of  the boundary term. For~this, we have to transform the torsion vector $T_{\mu}$. Before~that, and~for completeness, let us give an explicit formula for the contortion tensor. In~order to calculate it, we need to have the torsion tensor with all indices at the same level. This is easily done by considering  ${\tilde T}_{\alpha\mu\nu}={\tilde e}^b_{\alpha}\eta_{ab}{\tilde T}^a_{\hphantom{\beta}\mu\nu}$. We get
\begin{equation*}
    \begin{split}
{\tilde T}_{\alpha\mu\nu} & = C^2 T_{\alpha\mu\nu}+CD\phi_{\alpha}\phi^{\beta}T_{\beta\mu\nu}+C(C_{\mu}g_{\alpha\nu}-C_{\nu}g_{\alpha\mu})+\left((CD)_{\mu}\phi_{\nu}-(CD)_{\nu}\phi_{\mu}\right)\phi_{\alpha}  \\
& +D(\partial\phi)^2(D_{\mu}\phi_{\nu}-D_{\nu}\phi_{\mu})\phi_{\alpha}+CD(\phi_{\nu}\bigtriangledown_{\mu}\phi_{\alpha}-\phi_{\mu}\bigtriangledown_{\nu}\phi_{\alpha})+D^2\phi_{\alpha}\phi^{\beta}(\phi_{\nu}\bigtriangledown_{\mu}\phi_{\beta}-\phi_{\mu}\bigtriangledown_{\nu}\phi_{\beta}),
    \end{split}
\end{equation*}
and then, the contortion tensor $K_{\alpha\mu\nu}=\frac12( T_{\alpha\mu\nu}+ T_{\nu\alpha\mu} +T_{\mu\alpha\nu})$, antisymmetric in the lateral indices $\alpha-\nu$, is given by
\begin{equation}
\begin{split}
{\tilde K}_{\alpha\mu\nu} & = C^2 K_{\alpha\mu\nu}+CD\phi^{\beta}\left(\phi_{\mu}T_{\beta\alpha\nu}+\frac12 (\phi_{\alpha}T_{\beta\mu\nu}-\phi_{\nu}T_{\beta\mu\alpha})\right)\\
& +\frac12 CD\left(\vphantom{\frac12}\phi_{\nu}(\bigtriangledown_{\mu}\phi_{\alpha}+\bigtriangledown_{\alpha}\phi_{\mu})-\phi_{\alpha}(\bigtriangledown_{\mu}\phi_{\nu}+\bigtriangledown_{\nu}\phi_{\mu})\right)-D^2 \phi_{\mu}\phi^{\beta}(\phi_{\alpha}\bigtriangledown_{\nu}\phi_{\beta}-\phi_{\nu}\bigtriangledown_{\alpha}\phi_{\beta})\\
& +C(C_{\alpha}g_{\mu\nu}-C_{\nu}g_{\mu\alpha})-\phi_{\mu}\left((CD)_{\nu}\phi_{\alpha}-(CD)_{\alpha}\phi_{\nu}\right)+D(\partial\phi)^2\phi_{\mu}(D_{\alpha}\phi_{\nu}-D_{\nu}\phi_{\alpha}).
\end{split}
\end{equation}

Another very important quantity is the vector part of the disformally transformed torsion tensor $T_{\mu}=T^{\rho}_{\hphantom{\rho}\mu\rho}$, and~it can be obtained just by contracting the correct indices in (\ref{disftors}):
\begin{equation}
\begin{split}
{\tilde T}_{\mu} & = T_{\mu} - \frac{D}{C+D(\partial\phi)^2}\phi^{\beta}\phi^{\alpha}T_{\beta\mu\alpha} + 3\frac{C_{\mu}}{C}+\frac{(\partial\phi)^2(CD_{\mu}-DC_{\mu})}{C(C+D(\partial\phi)^2)} + \frac{D}{C+D(\partial\phi)^2}\phi_{\alpha}\bigtriangledown_{\mu}\phi^{\alpha}\\
& +\left(\frac{DC_{\alpha}\phi^{\alpha}-CD_{\alpha}\phi^{\alpha}}{C(C+D(\partial\phi)^2)} - \frac{D}{C}\Box \phi+ \frac{D^2}{C(C+D(\partial\phi)^2)}\phi^{\alpha}\phi^{\beta}\bigtriangledown_{\alpha}\phi_{\beta}\right)\phi_{\mu}.   
\end{split}
\end{equation}

\section{Disformal transformations in f(T) gravity\\ and the search for the Einstein frame}
\label{sec:disformalfT}

Let us recall that our main goal was to search for an Einstein frame, i.e.,~a frame in which the action will be written in terms of usual gravity with minimally coupled matter, and~in particular, without local Lorentz-violating terms. Since we assume that the disformal transformation depends only on the scalar field $\phi$, and~does not depend on any Lorentz-breaking quantities, then, according to the relation \eqref{equiv}, instead of transforming the torsion scalar, it is enough to look at the transformation of the Levi--Civita divergence of the torsion~vector.

Let us substitute the Equation \eqref{equiv} into the Jordan frame action \eqref{Jfr}:
\begin{equation}
S = \dfrac{1}{2 \kappa} \int d^4 x\ E \ \left(-f^{\prime}(\phi)\left(R(\mathop\Gamma\limits^{(0)})+2 \mathop\bigtriangledown\limits^{(0)}{\vphantom{\bigtriangledown}}_{\mu}T^{\mu}\right) + f(\phi)-\phi f^{\prime}(\phi)\right),
\end{equation}
or into the Legendre transform version of it
\eqref{JfrL}:
\begin{equation}
S =- \dfrac{1}{2\kappa} \int d^4 x \ E \ \left(\phi \left(R(\mathop\Gamma\limits^{(0)})+2 \mathop\bigtriangledown\limits^{(0)}{\vphantom{\bigtriangledown}}_{\mu}T^{\mu}\right)+ V(\phi)\right).
\end{equation}

{Of course, those two are different by an obvious scalar field redefinition. (This action can also be interpreted as a relevant part of any scalar-torsion theory with a non-minimal coupling to the torsion~scalar.)}

The only Lorentz-violation term is now the one with $\mathop\bigtriangledown\limits^{(0)}{\vphantom{\bigtriangledown}}_{\mu}T^{\mu}$, and by~integrating it by parts, we reduce the Lorentz-breaking effects to the action term of the form
\begin{equation}
 2\int d^4 x\ E\ f^{\prime\prime}(\phi)T_{\mu}\partial^{\mu}\phi \qquad {\mathrm or} \qquad 2\int d^4 x\ E\ T_{\mu}\partial^{\mu}\phi,
\end{equation}
depending on the chosen~representation.

The question was whether it is possible to get rid of this term by a disformal transformation in terms of the same scalar field which enters the Jordan frame. We have already computed all the necessary ingredients for that. The~calculation gives a {\it remarkably simple} result
\begin{equation}
\tilde{\phi^{\mu}}{\tilde T}_{\mu}=\frac{1}{(C+D(\partial\phi)^2)^2}\left(\phi^{\mu}T_{\mu}+3\frac{C_{\mu}}{C}\phi^{\mu}+\frac{D}{C}\left(\phi^{\alpha}\phi^{\beta}\bigtriangledown_{\alpha}\phi_{\beta}-(\partial\phi)^2\square\phi\right)\right)
\label{preTphi}
\end{equation}
where we have used that  $\phi^{\alpha}\phi^{\mu}\phi^{\beta}T_{\beta\mu\alpha}=0$ due to antisymmetry of the torsion tensor. Note that the latter expression would not have vanished if an independent scalar field was used for the transformation, leaving us with even more Lorentz-violating~terms.

We still need to improve this expression to make it more comprehensible. Indeed, the~covariant derivatives in \eqref{preTphi} are given in terms of the Weitzenb\"{o}ck connection, and~in order to separate the Lorentz-violating effects, we must reduce them to the Levi--Civita ones. This is easily done. One~expression does not require anything:
\begin{equation}
\phi^{\alpha}\phi^{\beta}\bigtriangledown_{\alpha}\phi_{\beta}=\phi^{\alpha}\phi^{\beta}( \mathop\bigtriangledown\limits^{(0)}{\vphantom{\bigtriangledown}}_{\alpha}\phi_{\beta}-K_{\mu\alpha\beta}\phi^{\mu})=\phi^{\alpha}\phi^{\beta} \mathop\bigtriangledown\limits^{(0)}{\vphantom{\bigtriangledown}}_{\alpha}\phi_{\beta}
\end{equation}
where we have used antisymmetry of the contortion tensor, and~for the d'Alembertian operator we get:
\begin{equation}
\square\phi=\mathop\square\limits^{(0)}\phi+K^{\mu}_{\hphantom{\mu}\mu\alpha}\phi^{\alpha}=\mathop\square\limits^{(0)}\phi-\phi^{\mu}T_{\mu}.    
\end{equation}

Mixing all these ingredients correctly, we get a pleasing result for the disformally transformed Lorentz-violating term
\begin{equation}
\label{themain}
\tilde{\phi^{\mu}}{\tilde T}_{\mu}=\frac{1}{(C+D(\partial\phi)^2)^2}\left[\left(1+\frac{D}{C}(\partial\phi)^2\right)\phi^{\mu}T_{\mu}+3\frac{C_{\mu}}{C}\phi^{\mu}+\frac{D}{C}\left(\phi^{\alpha}\phi^{\beta}\mathop\bigtriangledown\limits^{(0)}{\vphantom{\bigtriangledown}}_{\alpha}\phi_{\beta}-(\partial\phi)^2\mathop\square\limits^{(0)}\phi\right)\right].    
\end{equation}

If we recall now that $ {\tilde E} =C^3(C+D(\partial\phi)^2) E $, then we readily see that the coefficient in front of the $ E  T_{\mu}\partial^{\mu}\phi$ term changes simply by the factor of $C^2$:
$${\tilde E} {\tilde T}_{\mu}{\tilde\partial}^{\mu}\phi= E C^2 T_{\mu}\partial^{\mu}\phi+\ {\mathrm invariant\ terms}.$$

{Therefore, it is not possible to remove the Lorentz-violating term by means of disformal transformations.}

\section{Discussion and conclusions}
\label{sec:concl}

We have presented explicit formulas for disformal transformations of the basic quantities in teleparallel theories of gravity, mostly having in mind the $f(\mathbb T)$ theory. They are not able to bring the theory to a genuine Einstein frame~form.

An obvious caveat is that we could also allow for Lorentz-violating arguments in $C$ and $D$. However, as we have discussed above, it is highly unlikely that it will help eliminate Lorentz-breaking terms from the action. And, in~this case, it would not be enough for our purposes to study only the transformation of the boundary~term.

On the other hand, this Lorentz-violating term in the action vanishes whenever the torsion vector is orthogonal to the scalar gradient. One can think of that in terms of the fact that the Lorentz violation in $f(\mathbb T)$ gravity is related to the direction in which the torsion scalar, or~the auxiliary scalar field,~changes. Though, of~course, vanishing of an action term for a particular combination of fields does not have a very clear meaning since it is not respected by generic~variations.

Note, however, that after the disformal transformation, this vanishing can happen in other, more~interesting cases, too. Let us choose a very simple transformation with $C=D=1$. \mbox{For~it, the final} answer is even simpler:
\begin{equation}
\tilde{\phi}^{\mu} \tilde{T}_{\mu} = \dfrac{1}{1+(\partial\phi)^2}\left((1+(\partial\phi)^2)\phi^{\mu}T_{\mu} + \phi^{\mu}\phi^{\alpha} \mathop\bigtriangledown\limits^{(0)}{\vphantom{\bigtriangledown}}_{\mu}\phi_{\alpha}-(\partial\phi)^2\mathop\square\limits^{(0)}\phi  \right).
\end{equation}

This term can be set to zero as a whole if the torsion vector is related to the scalar field in a very particular way, namely:
\begin{equation}
T_{\mu} = \dfrac{\phi_{\mu}\mathop\square\limits^{(0)}\phi - \phi^{\alpha} \mathop\bigtriangledown\limits^{(0)}{\vphantom{\bigtriangledown}}_{\mu}\phi_{\alpha} }{1+(\partial\phi)^2 }.
\label{vectordisf}
\end{equation}

What is a possible physical meaning of this result? We do not know. The~formula \eqref{vectordisf} relates Lorentz-breaking and Lorentz-preserving quantities in it, and, therefore, is not in line with the whole procedure we followed. However, it looks intriguing. Can it be related somehow to the cases when linear  perturbative analyses show no new propagating modes? Or to the remnant symmetry?

Another related idea would be to use a two-field disformal transformation of the form
\begin{equation}
{\tilde E}^a_{\mu}=E^a_{\mu}+\psi\cdot \phi^a \phi_{\mu}
\end{equation}
where $\psi$ is an independent field at our disposal. All the calculations above apply to this case just by setting $C=1$ and $D=\psi$. Then, we see that the violating term \eqref{themain} gets transformed to zero if
\begin{equation}
\psi=-\frac{\phi^{\nu}T_{\nu}}{(\partial\phi)^2\phi^{\mu}T_{\mu} + \phi^{\mu}\phi^{\alpha} \mathop\bigtriangledown\limits^{(0)}{\vphantom{\bigtriangledown}}_{\mu}\phi_{\alpha}-(\partial\phi)^2\mathop\square\limits^{(0)}\phi}.
\end{equation}

Of course, the~rest of the Lagrangian receives then Lorentz-violation through the $\psi$ field, but~formally, it looks like a metric theory with two scalar fields, $\phi$ and $\psi$. Can we now treat these just as two independent scalar fields?

Whatever the answers are, we think that it shows that disformal transformations have good potential to shed some new light on $f(\mathbb T)$ gravity models. Furthermore,~since the latter are currently so important in cosmological model building, it is worthwhile to further pursue this~topic.

\section*{Acknowledgments}

M. J. G. was funded by CONICYT-FONDECYT postdoctoral grant 3190531. M. J. G. would like to thank the hospitality of colleagues at the Centre for Theoretical Physics of the British University in Egypt, where part of this work was completed. A.G. would like to thank the organizers of the 10th Mathematical Physics Meeting: School and Conference on Modern Mathematical Physics which was held on 9 - 14 September 2019 in Belgrade, Serbia.

\end{document}